\documentclass{svjour2}

\usepackage{graphicx}  
\usepackage{latexsym}
\usepackage{amssymb}

\def\beq{\begin{equation}}
\def\eeq{\end{equation}}

\journalname{General Relativity and Gravitation}

\begin{document}

\title{Hyperbolic-like elastic scattering of spinning particles by a Schwarzschild black hole}

\author{Donato Bini \and 
Andrea Geralico
}

\institute{Donato Bini 
\at
Istituto per le Applicazioni del Calcolo ``M. Picone,'' CNR, Via dei Taurini 19, I-00185, Rome, Italy\\
ICRANet, Piazza della Repubblica 10, I-65122 Pescara, Italy\\
INFN, Sezione di Napoli,
Via Cintia, Edificio 6 - 80126 Napoli, Italy \\
\email{donato.bini@gmail.com} 
\and
Andrea Geralico 
\at
Istituto per le Applicazioni del Calcolo ``M. Picone,'' CNR, Via dei Taurini 19, I-00185, Rome, Italy\\
ICRANet, Piazza della Repubblica 10, I-65122 Pescara, Italy\\
\email{geralico@icra.it}       
}

\date{Received: date / Accepted: date / Version: \today}

\maketitle

\begin{abstract}
The scattering of spinning test particles by a Schwarzschild black hole is studied.
The motion is described according to the Mathisson-Papa\-petrou-Dixon model for extended bodies in a given gravitational background field.
The equatorial plane is taken as the orbital plane, the spin vector being orthogonal to it with constant magnitude.
The equations of motion are solved analytically in closed form to first-order in spin and the solution is used to compute 
corrections to the standard geodesic scattering angle as well as capture cross section by the black hole.
\end{abstract}

\section{Introduction}

After the discovery of the gravitational waves in 2015 \cite{Abbott:2016blz} a new window in gravitational physics was opened, looking for any suitable astrophysical situation which could lead to the detection of new events. Binary systems in this sense play a central role. Indeed, the first detection has concerned a binary black hole system, whose evolution has been followed along all inspiral-coalescing-merging-plunging phases. 
Another promising situation is that of hyperbolic encounters or scattering by two black holes.
This problem  is, in principle, as interesting as the previous one, but its description is more complicated. For example, an almost circularized binary system emits gravitational waves at the frequency of its orbital motion and hence has its spectrum very compact around that value. The spectrum associated with a hyperbolic scattering of two black holes, instead, covers a large range of frequencies, and the theoretical predictions in terms of observed flux have not been developed much beyond the pioneering works of the 70's~\cite{rr-wheeler,Rees:1974iy}, so that most of the existing analysis belongs to lowest-order Post-Newtonian approximation \cite{Blanchet:1989cu} and  numerical relativity \cite{Pretorius:2007jn,Shibata:2008rq,Sperhake:2008ga,Sperhake:2009jz,Sperhake:2012me}. Semi-analytical methods, like the Effective-One-Body formalism, are trying to fill the gap, but no results beyond the 3PN approximation level have been shown in the literature up to now \cite{Majar:2010em,Bini:2012ji,Damour:2014afa,DeVittori:2014psa}.

We study here the problem of scattering of a spinning test particle by a Schwarzschild black hole according to the Mathisson-Papapetrou-Dixon (MPD) model \cite{Mathisson:1937zz,Papapetrou:1951pa,Dixon:1970zza}, in comparison with the well known case of spinless particles moving along hyperbolic-like geodesic orbits.
The equatorial plane is chosen as the orbital plane, implying that the spin vector is orthogonal to it with constant magnitude, as a consequence of the MPD equations.
Taking advantage of the condition of \lq\lq small spin'' implicit in the MPD model to avoid backreaction effects on the background metric as well as of the existence of conserved quantities related to Killing symmetries, we are able to analytically solve the equations of motion in terms of Elliptic integrals.
We compute the corrections to first-order in spin to the scattering angle, i.e., the most natural gauge-invariant and physical observable associated with the scattering process. 
Using a conical-like parametrization for the radial motion we express the scattering angle in terms of eccentricity and (inverse) semi-latus rectum, the latter parameters being in a $1$-$1$ correspondence with the two other natural variables: energy and angular momentum (which are also gauge invariant variables). 
We also determine the modification due to spin to the capture cross section by the black hole.
Finally, we compare the nongeodesic motion of a spinning particle discussed here with the companion situation of geodesic motion of a spinless particle orbiting a slowly rotating Kerr black hole in the weak field limit and discuss some reciprocity relations.

We will use geometrical units and conventionally assume that Greek indices run from $0$ to $3$, whereas Latin indices run from $1$ to $3$.

\section{Equatorial plane motion in a Schwarzschild spacetime}

The motion of a spinning test particle in a given gravitational background is described by the MPD equations \cite{Mathisson:1937zz,Papapetrou:1951pa,Dixon:1970zza}
\begin{eqnarray}
\label{papcoreqs1}
\frac{ DP^{\mu}}{d \tau} & = &
- \frac12 \, R^\mu{}_{\nu \alpha \beta} \, U^\nu \, S^{\alpha \beta}
\equiv  F^\mu_{\rm (spin)}\,,
\\
\label{papcoreqs2}
\frac{ DS^{\mu\nu}}{d \tau} & = & 
2 \, P^{[\mu}U^{\nu]} \,,
\end{eqnarray}
where $P^{\mu}\equiv m u^\mu$ (with $u \cdot u \equiv u^\mu u_\mu = -1$) is the total 4-momentum of the body with mass $m$ and unit direction $u^\mu$, $S^{\mu\nu}$ is the (antisymmetric) spin tensor, and $U^\mu=d x^\mu/d\tau$ is the timelike unit 4-velocity vector tangent to the body ``center-of-mass line,'' parametrized by the proper time $\tau$ (with parametric equations $x^\mu=x^\mu(\tau)$), used to make the multipole reduction.

In order to ensure that the model is mathematically self-consistent, the reference world line in the object should be specified by imposing some additional conditions. Here we shall use the Tulczyjew-Dixon conditions~\cite{Dixon:1970zza,tulc59}, which read
\beq
\label{tulczconds}
S^{\mu\nu}u{}_\nu=0\,. 
\eeq
With this choice, the spin tensor can be fully represented by a spatial vector (with respect to $u$),
\beq
S(u)^\alpha=\frac12
\eta(u)^\alpha{}_{\beta\gamma}S^{\beta\gamma}
\,,
\eeq 
where $\eta(u)_{\alpha\beta\gamma}=\eta_{\mu\alpha\beta\gamma}u^\mu$ is the
spatial unit volume 3-form (with respect to $u$) built from the unit volume 4-form
$\eta_{\alpha\beta\gamma\delta}=\sqrt{-g}\, \epsilon_{\alpha\beta\gamma\delta}$,
with $\epsilon_{\alpha\beta\gamma\delta}$ ($\epsilon_{0123}=1$) being the
Levi-Civita alternating symbol and $g$ the determinant of the metric. 

It is also useful to introduce the signed
magnitude $s$ of the spin vector
\beq
\label{sinv}
s^2=S(u)^\beta S(u)_\beta = \frac12 S_{\mu\nu}S^{\mu\nu}
\,,
\eeq
which is a constant of motion.
Implicit in the MPD model is the condition that the length scale $|s|/m$ naturally associated with the spin should be very small compared to the one associated with the background curvature (say $M$), in order to neglect back reaction effects, namely $|\hat s|\equiv {|s|}/{mM}\ll 1$.
Introducing this smallness condition from the very beginning leads to a simplified set of linearized differential equations around the geodesic motion. 
In fact, the total 4-momentum $P$ of the particle is aligned with $U$ in this limit, i.e., $P^{\mu}= mU^\mu+O(2)$, the mass $m$ of the particle remaining constant along the path. Furthermore, Eq. (\ref{papcoreqs2}) becomes ${ DS^{\mu\nu}}/{d \tau}=O(2)$, implying that the spin vector is parallely transported along $U$. 
 
Let us consider a spinning test particle moving in a Schwarzschild spacetime, whose metric written in standard spherical-like coordinates $(t,r,\theta,\phi)$ is
\beq
ds^2 =-N^2 dt^2+N^{-2}dr^2+r^2 (\sin^2\theta d\theta^2+d\phi^2)\,,
\eeq
where $N=(1-2M/r)^{1/2}$ denotes the \lq\lq lapse" function.
We introduce a family of fiducial observers, the static observers with 4-velocity 
\beq
n=N^{-1}\partial_t\,,
\eeq
equipped with an adapted triad
\beq
e_{\hat r}=N\partial_r \,,\quad e_{\hat \theta}=\frac{1}{r}\partial_\theta
\,,\quad e_{\hat \phi}=\frac{1}{r \sin \theta}\partial_\phi\,.
\eeq
We will limit our analysis to the case of equatorial motion, i.e., $\theta=\pi/2$ is the orbital plane and hence $U^\theta=0$.
As a convention, the physical (orthonormal) component along $-\partial_\theta$ which is perpendicular to the equatorial plane will be referred to as ``along the positive $z$-axis" and will be indicated by the index $\hat z$, when convenient: $e_{\hat z}=-e_{\hat \theta}$. 
From the evolution equations for the spin tensor it follows that the spin vector has a single nonvanishing and constant component along $\theta$ (or $z$), namely
\beq
S(U)=S^{\hat\theta}e_{\hat \theta}=s\, e_{\hat z}\,, \qquad
S^{\hat\theta}=-s\,. 
\eeq

Let us decompose the 4-velocity $U$ of the spinning particle with respect to the static observers
\beq
U=\gamma(U,n) [n+ \nu(U,n)]\,,\quad 
\nu(U,n)\equiv \nu(U,n)^{\hat r}e_{\hat r}+\nu(U,n)^{\hat \phi}e_{\hat \phi}\,,
\eeq
where $\gamma(U,n)=1/\sqrt{1-||\nu(U,n)||^2}$ is the associated Lorentz factor.
Hereafter we will use the abbreviated notation $\gamma(U,n)\equiv\gamma$ and $\nu(U,n)^{\hat a}\equiv\nu^{\hat a}$.
The relation with the coordinate components of $U$ is
\beq
\label{Ucoord_comp}
U^t\equiv \frac{d t}{d \tau}=\frac{\gamma}{N}\,, \qquad 
U^r\equiv \frac{d r}{d \tau}=\gamma N\nu^{\hat r}\,, \qquad 
U^\phi\equiv \frac{d \phi}{d \tau}=\frac{\gamma \nu^{\hat \phi}}{r}\,.
\eeq
The spin force turns out to be
\beq
\label{Fspin}
F_{\rm (spin)}=\frac{3mM^2}{r^3}{\hat s}\gamma^2\nu^{\hat \phi}[\nu^{\hat r}n+e_{\hat r}]\,.
\eeq
The equations of motion (\ref{papcoreqs1}) then imply
\begin{eqnarray}
\frac{d\nu^{\hat r}}{d \tau}&=&
\frac{N^2-1}{2r\gamma N}+\gamma(\nu^{\hat \phi})^2\frac{3N^2-1}{2rN}
-\frac32M{\hat s}\frac{N^2-1}{\gamma r^2}\nu^{\hat \phi}[1+\gamma^2(\nu^{\hat \phi})^2]
\,,\nonumber\\
\frac{d \nu^{\hat \phi}}{d \tau}&=&
-\frac1{2r^2N}\gamma\nu^{\hat r}\nu^{\hat \phi}\left[
r(3N^2-1)-3M{\hat s}N(N^2-1)\nu^{\hat \phi}
\right]\,.
\end{eqnarray}
In order to solve these equations we take advantage of the existence of conserved quantities along the motion in stationary and axisymmetric spacetimes endowed with Killing symmetries, i.e., the energy $E$ and the total angular momentum $J$ associated with the timelike Killing vector $\xi=\partial_t$
and the azimuthal Killing vector $\eta=\partial_\phi$, respectively. 
They are given by
\begin{eqnarray}
\label{totalenergy}
E&=&-\xi_\alpha P^\alpha +\frac12 S^{\alpha\beta}\nabla_\beta \xi_\alpha\,,\nonumber\\
J&=&\eta_\alpha P^\alpha -\frac12 S^{\alpha\beta}\nabla_\beta \eta_\alpha\,,
\end{eqnarray}
where $\nabla_\beta \xi_\alpha=g_{t[\alpha,\beta]}$ and $\nabla_\beta \eta_\alpha=g_{\phi[\alpha,\beta]}$.
We then find
\beq
\label{EandJ}
\hat E=N\gamma\left[1+\hat s\frac{M^2}{r^2N}\nu^{\hat \phi}\right]
\,,\qquad
\hat J=\frac{r}{M}\gamma\left[\nu^{\hat \phi}+\hat sN\frac{M}{r}\right]
\,,
\eeq
where $\hat E\equiv E/m$ and $\hat J\equiv J/mM$ are dimensionless.
Eq. (\ref{EandJ}) thus provide two algebraic relations for the frame components $\nu^{\hat r}$ and $\nu^{\hat \phi}$ of the linear velocity, which once inserted in Eq. (\ref{Ucoord_comp}) finally yield
\begin{eqnarray}
\label{finaleqs}
\frac{d t}{d \tau}&=&
\frac{\hat E}{N^2}-\frac{M^3\hat J}{N^2r^3}\hat s
\,, \nonumber\\ 
\left(\frac{d r}{d \tau}\right)^2&=&
{\hat E}^2-N^2\left(1+\frac{{\hat J}^2}{r^2}\right)
+\frac{2M^2\hat E\hat J(r-3M)}{r^3}\hat s
\,, \nonumber\\ 
\frac{d\phi}{d \tau}&=&
\frac{M\hat J}{r^2}-\frac{M\hat E}{r^2}\hat s
\,,
\end{eqnarray}
to first-order in spin.

\section{The scattering angle: spin corrections to the standard geodesic value}

For the orbit of the spinning particle we assume a conical-like representation of the radial variable, i.e.,
\beq
\label{conica}
r=\frac{Mp}{1+e\cos \chi}\,,
\eeq
where both the semi-latus rectum $p$ and the  eccentricity $e\geq0$ are dimensionless parameters \cite{Chandrasekhar:1985kt}.
Bound orbits have $0\leq e<1$ and $0<\hat E<1$ and oscillate between a minimum radius $r_{\rm(per)}$ (periastron, $\chi=0$) and a maximum radius $r_{\rm(apo)}$ (apastron, $\chi=\pi$)
\beq
\label{peri_apo}
r_{\rm(per)}=\frac{Mp}{1+e}\,,\qquad
r_{\rm(apo)}=\frac{Mp}{1-e}\,,
\eeq
corresponding to the extremal points of the radial motion, i.e., 
\beq
\frac{dr}{d\tau}\Big|_{r=r_{\rm(per)}}=0=\frac{dr}{d\tau}\Big|_{r=r_{\rm(apo)}}\,.
\eeq
These conditions can be used to express $\hat E$ and $\hat J$ in terms of $(p,e)$ as follows
\begin{eqnarray}
\label{en_ang_mom}
\hat E &=&\hat E_0+\hat s\hat E_{\hat s}
= \sqrt{\frac{(p-2)^2-4e^2}{p(p-3-e^2)}} - \hat s \frac{(1-e^2)^2}{2p(p-3-e^2)^{3/2} }
\,, \\
\hat J &=&\hat J_0+\hat s\hat J_{\hat s}
= \frac{p}{(p-3-e^2)^{1/2}} + \hat s \frac{(2p-9-3e^2)\sqrt{(p-2)^2-4e^2}}{2p^{1/2}(p-3-e^2)^{3/2}}
\,. \nonumber
\end{eqnarray}

In this paper we are interested in unbound (hyperbolic-like) orbits, i.e., with eccentricity $e>1$ and energy parameter $\hat E>1$, starting far from the hole at radial infinity, reaching a minimum approach distance $r_{\rm(per)}$, and then coming back to radial infinity, corresponding to $\chi\in [-\chi_{\rm (max)}, \chi_{\rm (max)}]$, $\chi_{\rm (max)}=\arccos(-1/e)$ (see Eq. (\ref{conica})). It is well known that Eqs. (\ref{peri_apo})--(\ref{en_ang_mom}) can be formally used also in this case, but they imply that the apoastron does not exist anymore, in the sense that it corresponds to a negative value of the radial variable.
The radial equation can then be converted into an equation for $\chi$
\begin{eqnarray}
M\frac{d\chi}{d\tau}&=&  u_p^2\hat J_0(1+e\cos \chi )^2\sqrt{1-6u_p-2u_p e \cos \chi}\nonumber\\
&&
\times\left[1 + \frac12u_p^2\hat E_0\hat J_0{\hat s}\frac{3+2 e\cos \chi-e^2}{1-6u_p-2u_p e \cos \chi}\right]\,,  
\end{eqnarray}
to first order in $\hat s$, so that the azimuthal equation finally becomes
\begin{eqnarray}
\label{eqphidichi}
\frac{d\phi}{d\chi} 
&=& \frac{1}{\sqrt{1-6u_p-2eu_p\cos \chi}}  -\hat s \frac{\hat E_0u_p (e\cos\chi+3)}{\hat J_0  (1-6u_p-2eu_p\cos \chi)^{3/2}}\,,
\end{eqnarray}
where $u_p=1/p$.
The solution of this equation can be obtained analytically in terms of Elliptic functions as
\beq
\phi(\chi)=\phi_0(\chi)+\hat s \phi_{\hat s}(\chi)\,,
\eeq
where
\begin{eqnarray}
\phi_0(\chi)&=&
\frac{\kappa}{\sqrt{e u_p}}  \left[ K(\kappa)-F\left(\cos\frac{\chi}{2},\kappa\right)\right]
\,,\nonumber\\
\phi_{\hat s}(\chi)&=&\frac{\hat E_0}{2\hat J_0}\left\{
\phi_0(\chi)
-\frac{\kappa}{\sqrt{e u_p}(1-6u_p-2eu_p)}  \left[ E(\kappa)-E\left(\cos\frac{\chi}{2},\kappa\right)\right]\right.\nonumber\\
&&\left.
-\frac{\kappa^2\sin\chi}{(1-6u_p-2u_pe)(1-6u_p-2u_pe\cos\chi)^{1/2}}
\right\}
\,,
\end{eqnarray}
with $\phi(0)=0$ and 
\beq
\kappa= 2\sqrt{\frac{e u_p}{1-6u_p+2eu_p}}\,.
\eeq
Here $K(k)$ and $F(\sin\varphi,k)$ and $E(k)$ and $E(\sin\varphi,k)$ are the complete and incomplete elliptic integrals of the first kind and of the second kind defined by
\beq
K(k)=\int_0^{\frac{\pi}{2}}\frac{dx}{\sqrt{1-k^2\sin^2x}}\,,\qquad
F(\sin\varphi,k)=\int_0^{\varphi}\frac{dx}{\sqrt{1-k^2\sin^2x}}\,,
\eeq 
and 
\beq
E(k)=\int_0^{\frac{\pi}{2}}\sqrt{1-k^2\sin^2x}\,dx\,,\qquad
E(\sin\varphi,k)=\int_0^{\varphi}\sqrt{1-k^2\sin^2x}\,dx\,,
\eeq 
respectively.

Unbound orbits which are not captured by the black hole start at an infinite radius at the azimuthal angle $\phi=\phi(-\chi_{\rm (max)})$, 
the radius decreases to its periastron value at $\phi=0$ and then returns back to infinite value at $\phi=\phi(\chi_{\rm (max)})$, undergoing a total increment of $\Delta\phi =\phi(\chi_{\rm (max)})-\phi(-\chi_{\rm (max)})$. This scattering process is symmetric with respect to the minimum approach ($\chi=0$) in the case of a spinless particle, for which $\phi_0(-\chi_{\rm (max)})=-\phi_0(\chi_{\rm (max)})$, so that $\Delta\phi =2\phi(\chi_{\rm (max)})$, and the deflection angle from the original direction of the orbit is $\delta_0(u_p,e)=2\phi_0(\chi _{\rm(max)})-\pi$, i.e.,
\beq
\delta_0(u_p,e)= \frac{2\kappa}{\sqrt{e u_p}} \left[ K\left(\kappa \right)  
-  F \left(\sqrt{\frac{e-1}{2e}}
,\kappa \right)\right]
-\pi\,.
\eeq
This feature maintains also in the case of spinning particles, for which
\beq
\delta(u_p,e,\hat s)=\delta_0(u_p,e)+\hat s \delta_{\hat s}(u_p,e)\,,
\eeq
with
\begin{eqnarray}
\delta_{\hat s}(u_p,e)&=&2\phi_{\hat s}(\chi_{\rm (max)})
=\frac{\hat E_0}{\hat J_0}\frac{\kappa}{\sqrt{e u_p}}\left\{
K\left(\kappa \right) -  F \left(\sqrt{\frac{e-1}{2e}},\kappa \right)\right.\nonumber\\
&-&\left.
\frac{1}{1-6u_p-2eu_p}  \left[ 
E(\kappa)-E\left(\sqrt{\frac{e-1}{2e}},\kappa \right)
+\kappa\sqrt{\frac{u_p(e^2-1)}{e(1-4u_p)}}
\right]
\right\}
\,.\nonumber\\
\end{eqnarray}

\begin{figure}
\centering
\includegraphics[scale=0.4]{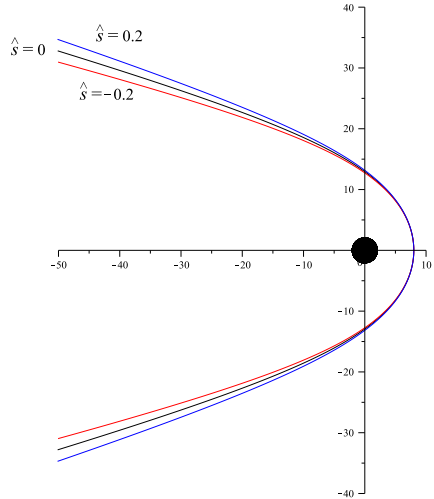}
\caption{The hyperbolic-like orbits of a spinning test particle with $\hat s=\pm0.2$ are shown in the $r$-$\phi$ plane for the choice of parameters $p=20$ and $e=1.5$ in comparison with the geodesic of a spinless particle ($\hat s=0$). 
The values of both energy and angular momentum per unit mass are different depending on $\hat s$, whereas the minimum approach distance $r_{\rm(per)}=8M$ is the same for all cases.
The deflection angle is $\delta\approx2.59122$ (i.e., $148.46608$ deg) for $\hat s=0$, $\delta\approx2.65601$ (i.e., $152.17820$ deg) for $\hat s=-0.2$, and $\delta\approx2.52643$ (i.e., $144.75397$ deg) for $\hat s=0.2$.
}
\label{fig:1}
\end{figure}

Figure \ref{fig:1} (a) shows a typical hyperbolic-like orbit of a spinning particle with spin aligned along the positive $z$-axis and in the opposite direction compared with the corresponding geodesic orbit of a spinless particle.
The orbital parameters are chosen as $p=20$ and $e=1.5$, implying that the distance of minimum approach is $r_{\rm(per)}=8M$.
The trajectory of the spinning particle thus depends on the same parameters $p$ and $e$ as in the spinless case.
Once these parameters have been fixed, orbits with different values of $\hat s$ have the same closest approach distance, but different values of energy and angular momentum.  
On the contrary, setting a pair of values of ($\hat E$, $\hat J$) leads to a shift of the periastron due to spin, as shown below.

                          
\subsection{Periastron shift}

A different (equivalent) parametrization of the orbit can be adopted in terms of energy $\hat E$ and angular momentum $\hat J$ instead of $p$ and $e$.
In this case the periastron distance depends on $\hat s$, and can be determined from the turning points for radial motion.
The equation of the orbit can be written as follows in terms of the dimensionless inverse radial variable $u = M/r$
\begin{eqnarray}
\label{equdiphi}
\left(\frac{d u}{d \phi}\right)^2&=&
\frac{2}{\hat J}({\hat J}-{\hat E}{\hat s})\left[
u^3-\frac{1}{2\hat J}({\hat J}+{\hat E}{\hat s})u^2+\frac{1}{{\hat J}^3}({\hat J}+3{\hat E}{\hat s})u\right.\nonumber\\
&&\left.
+\frac{1}{2{\hat J}^3}({\hat E}^2-1)({\hat J}+3{\hat E}{\hat s})
\right]\nonumber\\
&\equiv&
\frac{2}{\hat J}({\hat J}-{\hat E}{\hat s})(u-u_1)(u-u_2)(u-u_3)
\,.
\end{eqnarray}
For hyperbolic orbits we have $u_1 < 0 < u \leq u_2 < u_3$, with $u_2$ corresponding to the closest approach distance, i.e., $r_{\rm(per)}=M/u_2$ \cite{Chandrasekhar:1985kt}.
In the geodesic case ($\hat s=0$) these roots are given by
\begin{eqnarray}
u_{1\,(0)}&=& \frac16 \left[1-e^{-i\pi/3}X -e^{i\pi/3}\left(1-\frac{12}{{\hat J}^2}\right)\frac1{X} \right]\,,\nonumber\\
u_{2\,(0)}&=& \frac16 \left[1-e^{i\pi/3}X -e^{-i\pi/3}\left(1-\frac{12}{{\hat J}^2}\right)\frac1{X} \right]\,,\nonumber\\
u_{3\,(0)}&=& \frac16 \left[1+X+\left(1-\frac{12}{{\hat J}^2}\right)\frac1{X} \right]\,,
\end{eqnarray}
where
\beq
X=\frac{1}{\hat J}\left[\hat J\left({\hat J}^2-54{\hat E}^2+36\right) +6 i \sqrt{3}\sqrt{\Delta_{(0)}}\right]^{1/3}\,,
\eeq
When the discriminant 
\beq
\Delta_{(0)}=({\hat E}^2-1){\hat J}^4-(27{\hat E}^4-36{\hat E}^2+8){\hat J}^2-16
\eeq
is positive, the roots are all real, which is the case under consideration here. 
To first-order in spin, the corrections to the geodesic values are
\begin{eqnarray}
u_1&=&u_{1\,(0)}-\frac{\hat E}{2{\hat J}^3}\frac{3({\hat E}^2-1)+u_{1\,(0)}(6-u_{1\,(0)}{\hat J}^2)}{(u_{1\,(0)}-u_{3\,(0)})(u_{1\,(0)}-u_{2\,(0)})}\hat s
\,,\nonumber\\
u_2&=&u_{2\,(0)}-\frac{\hat E}{2{\hat J}^3}\frac{3({\hat E}^2-1)+u_{2\,(0)}(6-u_{2\,(0)}{\hat J}^2)}{(u_{2\,(0)}-u_{1\,(0)})(u_{2\,(0)}-u_{3\,(0)})}\hat s
\,,\nonumber\\
u_3&=&u_{3\,(0)}-\frac{\hat E}{2{\hat J}^3}\frac{3({\hat E}^2-1)+u_{3\,(0)}(6-u_{3\,(0)}{\hat J}^2)}{(u_{3\,(0)}-u_{1\,(0)})(u_{3\,(0)}-u_{2\,(0)})}\hat s
\,,
\end{eqnarray}
satisfying the properties
\beq
u_1+u_2+u_3=\frac12\left(1+\frac{\hat E}{\hat J}\hat s\right)\,.
\eeq
Note that $u_2$ and $u_3$ can be obtained from $u_1$ simply by replacing indices cyclically.
Eq. (\ref{equdiphi}) then yields
\beq
\label{eqphidiu}
\frac{d \phi}{d u}=\pm\frac1{\sqrt{2}}\left(1+\frac{{\hat E}}{2\hat J}{\hat s}\right)\frac1{\sqrt{(u-u_1)(u_2-u)(u_3-u)}}\,,
\eeq
where the $\pm$ sign should be chosen properly during the whole scattering process, depending on the choice of initial conditions.
For instance, by choosing $\phi(u_2)=0$ at periastron, integration between $u_2$ and $u$ gives
\beq
\phi(u)=\mp\frac{\sqrt{2}}{\sqrt{u_3-u_1}}\left(1+\frac{{\hat E}}{2\hat J}{\hat s}\right)
\left[K(m)-F(\alpha(u),m)\right]
\,,
\eeq
where
\beq
m=\sqrt{\frac{u_2-u_1}{u_3-u_1}}\,,\qquad
\alpha(u)=\sqrt{\frac{u-u_1}{u_2-u_1}}\,.
\eeq
The deflection angle is then 
\begin{eqnarray}
\delta(\hat E,\hat J,\hat s)
&=&\frac{2\sqrt{2}}{\sqrt{u_3-u_1}}\left(1+\frac{{\hat E}}{2\hat J}{\hat s}\right)
\left[K(m)-F(\alpha(0),m)\right]-\pi\,.
\end{eqnarray}

\begin{figure}
\centering
\[\begin{array}{cc}
\includegraphics[scale=0.3]{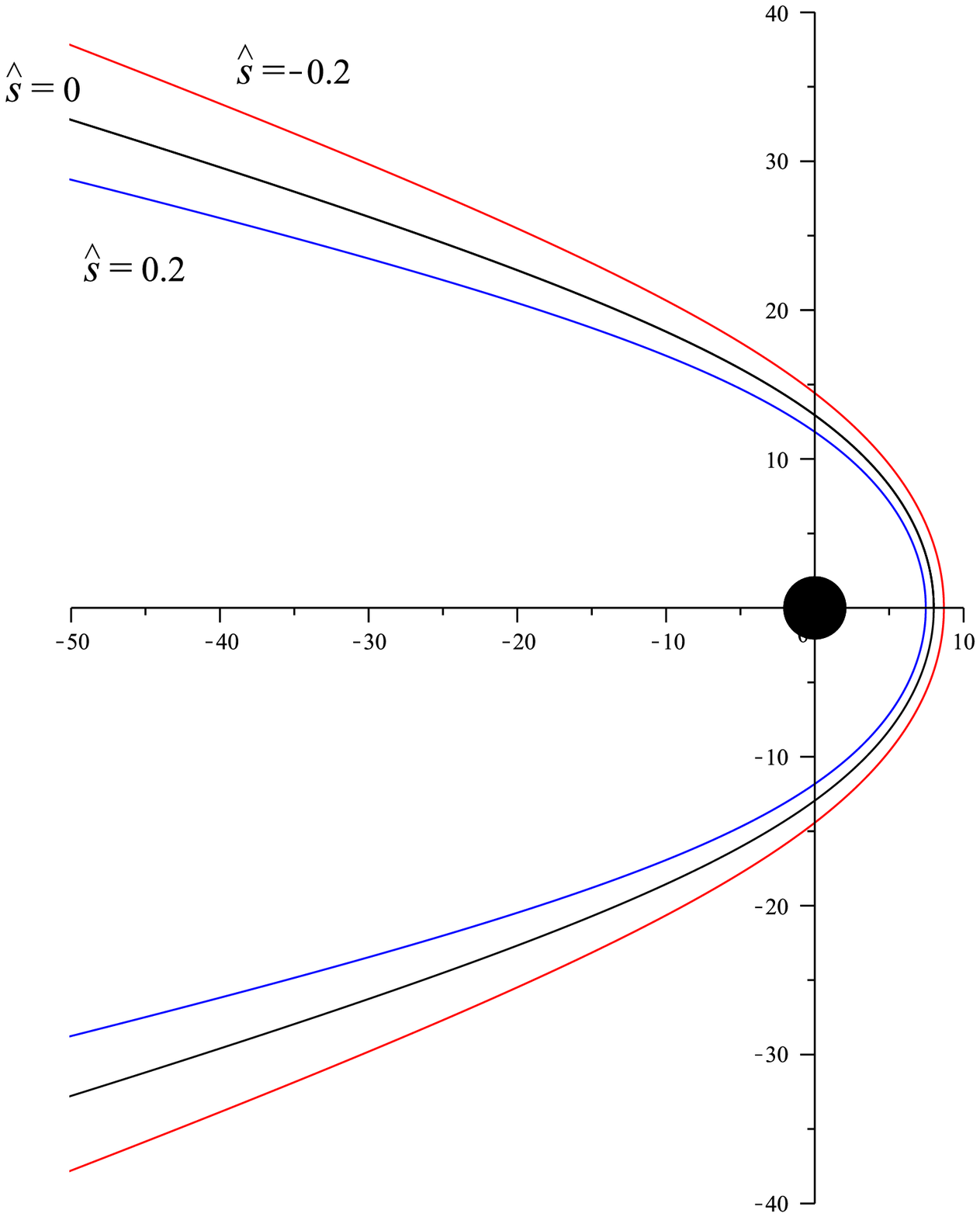}&
\includegraphics[scale=0.3]{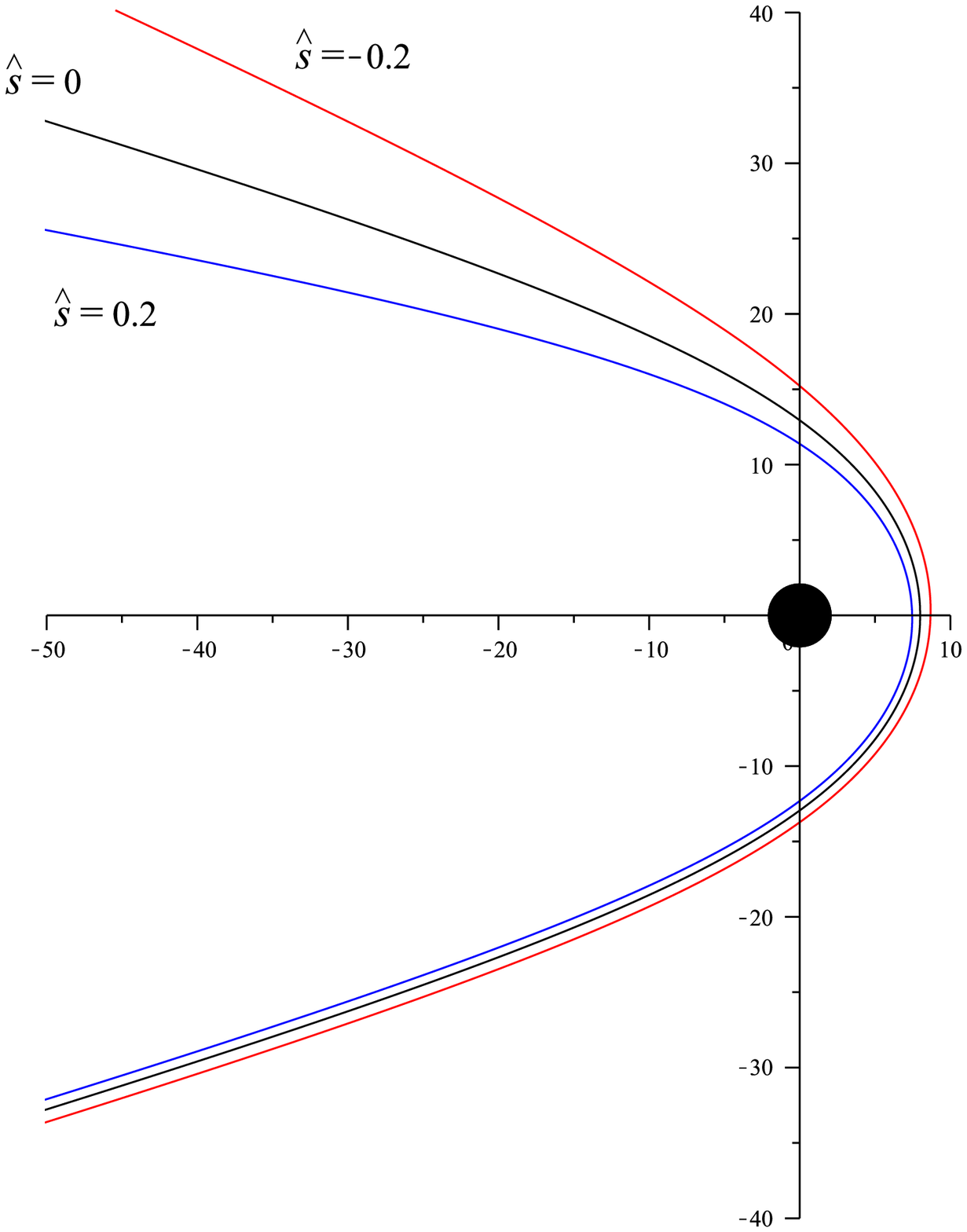}\\
(a)&(b)\\
\end{array}
\]
\caption{The hyperbolic-like orbits of a spinning test particle with $\hat s=\pm 0.2$ are shown in the $r$-$\phi$ plane for the choice of parameters $\hat E\approx1.03334$ and $\hat J\approx5.20756$ for two different sets of initial conditions.
The reference geodesic is the same as in Fig. \ref{fig:1}.
In panel (a) the orbital equation (\ref{eqphidiu}) has been integrated starting from the periastron $r_{\rm(per)}=M/u_2$ located at $\phi(u_2)=0$ for every value of $\hat s$. 
The deflection angle is $\delta\approx2.59122$ (i.e., $148.46608$ deg) for $\hat s=0$, $\delta\approx2.45760$ (i.e., $140.80996$ deg) for $\hat s=-0.2$, and $\delta\approx2.70531$ (i.e., $155.00262$ deg) for $\hat s=0.2$.
In panel (b), instead, the initial conditions have been chosen in such a way that all trajectories start from a common spacetime point far from the hole ($u=0$, $\phi(0)\approx-2.86621$).
The periastron is at  $u_2\approx0.11520$ (i.e., $r_{\rm(per)}\approx8.68044M$) and $\phi(u_2)\approx-0.06687$ for $\hat s=-0.2$, 
$u_2=0.125$ (i.e., $r_{\rm(per)}=8 M$) and $\phi(u_2)=0$ for $\hat s=0$
 and $u_2\approx0.13400$ (i.e., $r_{\rm(per)}\approx7.46247M$) and $\phi(u_2)\approx0.05704$ for $\hat s=0.2$.
}
\label{fig:2}
\end{figure}

A typical orbit using this parametrization is shown in Fig. \ref{fig:2} (a). 
The two branches (upper $+$, lower $-$) join at the periastron on the horizontal axis. 
In Fig. \ref{fig:2} (b), instead, initial conditions are such that all trajectories start from a common spacetime point far from the hole ($u=0$, $\phi(0)=\phi_-$), and complete their scattering process with different minimum approach distances $r_{\rm(per)}=M/u_2$ at $\phi=\phi(u_2)$ depending on the value of the spin parameter.

                          
\subsection{Capture cross section}

The condition for capture by the black hole is $u_2 = u_3$, i.e., the cubic at the right hand side of Eq. (\ref{equdiphi}) has a double root, corresponding to the vanishing of the discriminant \cite{Metner:1963,Misner:1974qy,Collins:1973xf}.
Solving for $\hat J$ then yields the critical value of the dimensionless angular momentum for capture
\beq
{\hat J}_{\rm crit}^2={\hat J}_{\rm crit\,(0)}^2\left[
1+\frac{2\hat s}{{\hat J}_{\rm crit\,(0)}\hat E(9{\hat E}^2-8)}
\left(3{\hat E}^2-2-\frac{8}{{\hat J}_{\rm crit\,(0)}^2}\right)
\right]\,,
\eeq
where
\beq
{\hat J}_{\rm crit\,(0)}^2=\frac{27{\hat E}^4-36{\hat E}^2+8+{\hat E}(9{\hat E}^2-8)^{3/2}}{2{\hat E}^2-1}\,.
\eeq
The associated cross section is thus given by 
\beq
\sigma_{\rm capt}=\pi b_{\rm crit}^2\,, \qquad
b_{\rm crit}=M\frac{{\hat J}_{\rm crit}}{\sqrt{{\hat E}^2-1}}\,,
\eeq
where $b_{\rm crit}$ is the critical impact parameter.
In the ultrarelativistic limit ($\hat E\gg1$) the first order correction in spin to the capture cross section is thus given by
\beq
\sigma_{\rm capt}\big|_{\rm u.r.}=27\pi M^2\left[1+\frac{2}{3\hat E^2}\left(1+\frac{\sqrt{3}}{9}\hat s\right)+O\left(\frac1{\hat E^4}\right)\right]\,,
\eeq
whereas for low energies ($\beta^2\equiv E^2-1\ll1$)
\beq
\sigma_{\rm capt}\big|_{\rm n.r.}=16\pi M^2\frac1{\beta^2}\left(1+\frac14\hat s\right)+O(\beta^0)\,.
\eeq

\section{Discussion}

Most of the existing studies on spinning particle motion in a Schwarzschild spacetime concern either circular or eccentric orbits or even deviations due to spin from a reference orbit which is a circular geodesic (i.e., quasi-circular orbits) 	\cite{Rietdijk:1992tx,Apostolatos:1996,Bini:2004md,Bini:2005nt,Plyatsko:2005bh,Mohseni:2010rm,Bini:2014poa,Plyatsko:2011wp,Costa:2012cy,Harms:2016ctx}.
These results are useful when taking into account backreaction effects to the background metric, going beyond the test particle approximation by using, e.g., self-force techniques \cite{Sasaki:2003xr}.
Furthermore, it has been shown in Ref. \cite{Burko:2003rv} that the conservative effect on the orbital dynamics in the extreme mass ratio limit is typically dominated by the spin force (with respect to the conservative part of the local self-force), whereas the decay of the orbit is dominated by radiation reaction.
As a consequence, in the construction of the gravitational waveforms the contribution of the spin-orbit coupling may be much more important for astrophysical systems than that due to the self-force.

We have considered here hyperbolic-like orbits, extending previous results.
The analytical solution of the orbit as well as observational effects like the correction due to spin to the scattering angle and the shift of the periastron may be useful for modeling spin effects in scattering processes in the framework of perturbation theory.
We briefly discuss below some reciprocity relations concerning the companion problem of a particle without structure moving along a hyperbolic-like geodesic in a slowly rotating Kerr spacetime.
Finally, we compare the effects due to the spin force with those of a drag force studied elsewhere.

\subsection{Comparison with the hyperbolic-like geodesic motion in a slowly rotating Kerr spacetime}

One can compare Eq. (\ref{eqphidichi}) governing the evolution of $\phi$ as a function of $\chi$ during the scattering process in the case of spinning particle orbiting a Schwarzschild black hole (in the small spin approximation) with the analogous equation valid for a spinless particle moving along a hyperbolic-like geodesic orbit in a Kerr spacetime (in the approximation of small rotation).

To first-order in the rotation parameter $\hat a=a/M$ the dimensionless energy and angular momentum are given by
\begin{eqnarray}
\hat E &=&\hat E_0-\hat a \frac{(1-e^2)^2}{p(p-3-e^2)^{3/2} }
\,, \nonumber\\
\hat J &=&\hat J_0-\hat a \frac{(3+e^2)\sqrt{(p-2)^2-4e^2}}{p^{1/2}(p-3-e^2)^{3/2}}
\,,
\end{eqnarray}
whereas the orbital equation reads
\begin{eqnarray}
\frac{d\phi}{d\chi}\Big|_{\rm kerr,\,geo} 
&=& \frac{1}{\sqrt{1-6u_p-2eu_p\cos \chi}}\nonumber\\
&-&
4\hat a \frac{\hat E_0u_p}{\hat J_0}\frac1{(1-6u_p-2eu_p\cos \chi)^{3/2}(1-2u_p-2eu_p\cos \chi)}
\,.\nonumber\\
\end{eqnarray}
In the weak field limit (i.e., $u_p\ll1$) the two equations of the orbit of a spinless (geodesic) particle in a Kerr background and a spinning particle in a Schwarzschild spacetime become
\begin{eqnarray}
\frac{d\phi}{d\chi}\Big|_{\rm kerr,\,geo} 
-\frac{d\phi}{d\chi}\Big|_{\rm schw,\,geo} 
&=& 
 -4\hat au_p^{3/2}+O(u_p^{5/2})
\,,\nonumber\\
\frac{d\phi}{d\chi}\Big|_{\rm schw,\,spin} 
-\frac{d\phi}{d\chi}\Big|_{\rm schw,\,geo}
&=&  
 -(3+e\cos\chi)\hat su_p^{3/2}+O(u_p^{5/2})
\,,
\end{eqnarray}
respectively, where
\beq
\frac{d\phi}{d\chi}\Big|_{\rm schw,\,geo} 
= 1+(e\cos\chi+3)u_p+\frac32(e\cos\chi+3)^2u_p^2
+O(u_p^3)\,.
\eeq
It then follows that to the leading order in $u_p$ (i.e., neglecting also corrections due to eccentricity)
the two equations are mapped one into the other simply by replacing $3 \hat s \to  4 \hat a$, or equivalently
\beq
\frac{3}{2} \hat s \quad \to \quad  2\hat a\,.
\eeq
Let us denote by $m_1=m$ and $S_1=s$ the mass and spin of the particle and by $m_2=M$ ($m_1\ll m_2$) and $S_2=m_2 a$ those of the black hole.
Restoring then the mass factors, i.e., $\hat s={S_1}/{m_1 m_2}$ and $\hat a=S_2/m_2^2$, the above relation becomes
\beq
\frac{3}{2} \frac{S_1}{m_1 m_2}  \quad \to \quad  2\frac{S_2}{m_2^2}\,,
\eeq
that is
\beq
\label{comparison}
\frac{3}{2} \frac{m_2 }{m_1 } S_1 \quad \to \quad  2 S_2\,.
\eeq
In the discussion of a two-body systems with spins~\cite{Damour:2007nc}, two new spin variables are known to play a role, namely
\beq
S_*=\frac{m_2 }{m_1 } S_1+\frac{m_1 }{m_2 } S_2\,,\qquad S=S_1+S_2\,.
\eeq
Here, from the Schwarzschild point of view $S_2=0$ and the only surviving spin variable is $S_*=(m_2/m_1) S_1$. Similarly, from the Kerr point of view (where the considered particle is spinless and moves along a geodesic) the only surviving spin variable is $S=S_2$. This means that Eq. (\ref{comparison}) can be cast in the form
\beq
\frac32 S_*  \quad \to \quad  2 S\,,
\eeq 
in the approximation specified above in which these considerations hold.
It is now easy to recognize the gyrogravitomagnetic ratios introduced in Ref. \cite{Damour:1992qi}, i.e.,
\beq
g_{S_*}=\frac32 \,,\qquad g_S=2\,,
\eeq
at their leading order values.
Finally, Eq. (\ref{comparison}) becomes
\beq
g_{S_*} S_* \quad \to \quad g_{S} S\,,
\eeq
as expected.

\subsection{Comparison of effects due to the spin force with those of a drag force}

In a recent work \cite{Bini:2016ubc} we have investigated the situation in which the particles under consideration were (structureless) test particles, whose deviation from geodesic motion was due to an (external) drag force $F_{\rm (drag)}$, chosen so that its components in the plane of motion are proportional to the corresponding components of the $4$-velocity itself, i.e., $F_{\rm (drag)}^r \propto U^r$ and $F_{\rm (drag)}^\phi \propto U^\phi$, namely
\beq
F_{\rm (drag)}=F_{\rm (drag)}^t \partial_t -\lambda \left(U^r \partial_r+ U^\phi \partial_\phi\right)\,, 
\eeq
with $\lambda$ a dimensionless constant modeling the physics of the dragging.
This is the case of particles interacting with accreting flows also in the presence of external electromagnetic fields or plasma \cite{McCourt:2015dpa}.
The temporal component follows from the orthogonality condition of $F_{\rm (drag)}$ and $U$, i.e., $F_{\rm (drag)}\cdot U=0$, leading to
\beq
F_{\rm (drag)}=-\lambda\gamma\left\{\left[(\nu^{\hat r})^2+(\nu^{\hat \phi})^2\right]n+\nu^{\hat r}e_{\hat r}+\nu^{\hat \phi}e_{\hat \phi}\right\}\,.
\eeq
This drag force acts on the orbital plane like a viscous force, so that is has dissipative effects, leading to the loss of energy and angular momentum during the scattering process.
In contrast, the spin force (\ref{Fspin}) acts as a conservative force, the total energy and angular momentum being constants of motion, implying that the scattering process is perfectly symmetric with respect to the minimum approach distance, as we have shown above.
In both cases, as a common feature particles are either scattered or captured by the black hole.

There also exist other kinds of dragging, like that leading to the well known Poynting-Robertson effect \cite{PR}, where the presence of a superposed photon test field implies the existence of a critical radius at which the radiation pressure balances the gravitational attraction, allowing rings of matter to form (see, e.g., Refs. \cite{abram,ML,Bini:2008vk,Bini:2011zza,Bini:2014yca} for recent applications to different backgrounds of astrophysical interest). 
The interplay between spin and radiation forces has been discussed in Ref. \cite{Bini:2010xa} by analyzing the deviation from circular geodesic motion.
A temporal counterpart to the Poynting-Robertson effect has been considered very recently in Ref. \cite{Bini:2016tqz}, where a distribution of collisionless dust around the black hole is responsible for the drag instead of the radiation field. 
In both cases the friction force has the form 
\beq
\label{force_U}
F_{\rm (drag)}^\alpha=- \sigma P(U)^\alpha{}_\mu T^{\mu\nu}U_\nu\,,
\eeq
where $P(U)$ projects orthogonally to $U$ and $\sigma$ denotes the (constant) effective interaction cross section, built with either the stress-energy tensor $T^{\mu\nu}=T_{\rm (rad)}^{\mu\nu}$ associated with the photon field or that $T^{\mu\nu}=T_{\rm (dust)}^{\mu\nu}$ associated with the dust field.
The existence of equilibrium orbits may prevent particles moving on a scattering orbit from either falling into the hole or escaping to infinity.

In future works we will extend the above discussion to the more interesting situation of a Kerr spacetime, where the rotation of the hole plays an important role.

\begin{acknowledgements}
D.B. thanks Prof. T. Damour for useful discussions and the Italian INFN (Salerno)  for partial support. 
\end{acknowledgements}

\end{document}